# Magnetoelasticity-driven phase inversion of ultrafast spin precession in $Ni_xFe_{100-x}$ thin films

Yooleemi Shin[1†], Seongsoo Yoon[2†], Jung-Il Hong[2]*, Ji-Wan Kim[1]*

[1]Department of Physics, Kunsan National University, Kunsan 54150, South Korea

[2]Department of Physics and Chemistry, DGIST, Daegu 42988, South Korea

**Abstract**

We present strong evidences for the deterministic role of magnetoelasticity in ultrafast spin dynamics of ferromagnetic $Ni_xFe_{100-x}$ alloy films. Without a change in the crystal structure, we observed sudden π-phase inversion of the spin precession in the range of $x = 87.0 – 97.5$. In addition, it was found that the phase was continuously changed and reversed its sign by varying the pump fluence. These cannot be explained simply by temperature dependence of magnetocrystalline, demagnetizing, and Zeeman fields which have been conventionally considered so far in describing the spin dynamics. Through the temperature- and composition-dependent simulations adding the magnetoelastic field generated from the lattice thermal strain, we revealed that the conventional and magnetoelastic fields were competing around $x = 95.3$, where the spin dynamics showed the largest phase shift. For analytic understanding, we further show that the temperature-dependent interplay of the Curie temperature, saturation magnetization, and magnetostriction, which are demonstrated to be the most important macroscopic parameters, determines the ultrafast spin dynamics. Our extensive study emphasizes that magnetoelasticity is the key ingredient for fully understanding the driving mechanism of ultrafast spin dynamics.

†: Equal contribution to this work

*: Corresponding authors.

E-mail address: jihong@dgist.ac.kr, hwoarang@kunsan.ac.kr

## 1. Introduction

The control of the spin degree of freedom through the complete understanding of ultrafast phenomena is a central requisite for the development of high-speed spintronic devices [1-6]. Ultrafast spin dynamics, particularly spin precessions arising after thermal equilibrium among electron, lattice, and spin system, has been macroscopically understood and described by the time-dependent interplay of constituent effective magnetic fields. Hence, it is important to understand all types of effective fields to clarify the driving mechanism of spin dynamics. In typical ferromagnets such as Co, Fe, Ni, and their base alloys or multilayers, the total effective field is comprised of magnetocrystalline, demagnetizing, and Zeeman fields [7-12]. In ferri- and antiferromagnetic oxides, the helicity-dependent effective field [13-15] and helicity-independent field owing to heating [16-18] can also be incorporated into the magnetocrystalline anisotropy field. Additionally, for the similar type of oxide materials, terahertz pulses have been used to directly induce the spin precession acting as an external magnetic field [19-21]. The exchange field between two sublattices for ferrimagnetic materials was considered to reveal all-optical spin reversal [22,23]. Recently, current pulse-induced switching in heavy metal-ferromagnet heterostructures has emerged as a promising technology for scalable spintronic devices [24-27]. Here, the spin-orbit torque consisting of both damping-like and field-like terms is introduced in Landau-Lifshitz-Gilbert (LLG) equation.

The absorption of laser photons and Joule heating of the current pulse normally increase the temperature of the materials, causing the thermal expansion of the lattice. Therefore, the piezo-effect from a dimensional change as well as the thermo-effect from a temperature change are expected to influence the spin dynamics to a non-negligible extent. However, despite the nonzero magnetostriction of various oxides and ferromagnets, magnetoelasticity has not been properly considered for materials mentioned above. The magnetoelastic effect originated from lattice expansion in Co and Ni films is strong enough to overcome other effects due to the conventional origins (e.g., magnetocrystalline, demagnetizing, and Zeeman) as demonstrated in the recent work of Ref [28]. Nevertheless, a phase evolution in spin precession depending on the driving mechanism has not been addressed so far. A comprehensive and systematic study by the variation in magnetic parameters, including Curie temperature, magnetization, and magnetostriction should be further explored.

In this work, to vary magnetoelasticity, we fabricated $Ni_xFe_{100-x}$ thin films with selected atomic compositions to which the spin dynamics became sensitive. It was found that the phase of the spin precession was abruptly inverted as $x$ varies around a specific value and exhibited a significant transition continuously, changing its sign with the pump fluence. By carrying out the extensive simulations of spin precessions with an additional consideration of magnetoelastic energy in the temperature-dependent LLG equation, we were able to demonstrate that all experimental features could be analyzed by the competition between the time-dependent magnetoelastic effect and other effects from the conventional energy terms. Our results provide a new perspective for an integrated understanding of the physical mechanism behind ultrafast spin dynamics, revealing an important factor that should be exploited in the optical design of ultrafast spintronic devices.

## 2. Experimental

To investigate the competition between the conventional and magnetoelastic effects in spin precession, 250-nm-thick polycrystalline $Ni_xFe_{100-x}$ alloy films with atomic compositions $x = 87.0, 90.9, 92.3, 95.3$, and $97.5$ were deposited on sapphire substrates (0001) by magnetron co-sputtering. To prevent oxidation, the MgO(3 nm) capping layer was deposited on each film. The atomic composition of each film was determined by energy dispersive X-ray spectroscopy. The in-plane hysteresis for the saturation magnetization was measured using a vibrating sample magnetometer. A well-known interesting feature of NiFe alloy is that the magnetostriction coefficient $\lambda_s$ changes its sign from positive at $x < 80$ to negative at $x > 80$ [29]. Hence, our samples compositions are suitable for investigating the transition of how the magnetoelasticity transforms ultrafast spin dynamics.

To measure ultrafast spin dynamics of ferromagnetic materials, the time-resolved magneto-optical polar Kerr effect was employed. We used femtosecond laser pulses issued from a Ti:sapphire oscillator system with a repetition rate of 80 MHz, temporal width of 40 fs, and center wavelength of 800 nm. The electro-optic modulator decreased the pulse rate to 1 MHz, and the temporally dispersed pulse was compressed to 45 fs using negatively chirped mirrors. The pump pulse for the two-color pump-probe measurement, was generated by frequency doubling from a 500-μm-thick $BiB_3O_6$ (BiBO) crystal with a conversion efficiency of 25 %. The pump (400 nm) and probe (800 nm) pulses were focused on the thin films through an objective lens (N.A. = 0.4). The pump pulse was slightly defocused at the sample plane to make the spot larger than that of the probe beam. The reflected probe pulse from the samples was split into s- and p-polarized beams using a Wollaston prism and analyzed into the differential Kerr rotation $\Delta\theta(t)$. For entire measurements, an external magnetic field of $H_{ext} = 0.4$ T was applied at an angle of $\varphi = 27$ ° from the film normal direction (z-axis).

## 3. Results and discussion

*3.1 Time-resolved magneto-optical Kerr effect data depending on atomic composition and pump fluence*

Fig. 1a and 1b shows the differential polar Kerr rotation $\Delta\theta(t)$ (solid circles) for all $x$ excited at the lowest pump fluence $I_{P,low}$ = 0.21 mJ/cm$^2$ (Fig. 1a) and the highest pump fluence $I_{P,high}$ = 4.6 mJ/cm$^2$ (Fig. 1b). The solid lines on experimental data for $x$ = 97.5 denote the fit curves matched with the following damped sinusoidal function:

$$\Delta\theta(t) = A\exp(-t/\tau_1)\sin(2\pi ft - \delta_0) + B\exp(-t/\tau_2) + C, \qquad (1)$$

where $\tau_1$ is the relaxation time of the spin precession; $\tau_2$ is the decay time of the background signal; $A$ and $B$ are the amplitudes of the two exponential decay terms; $C$ is the offset constant; $f$ is the frequency; and $\delta_0$ is the phase of the spin precession. Fig. 1a (for $I_{P,low}$) shows that $\delta_0$ is reversed at $x$ = 92.3 ~ 95.3, where the spin precession is not definite. This suggests that the dominant mechanism driving the spin precession changes. The spin precessions in Fig. 1b (for $I_{P,high}$) differ from the curves for $I_{P,low}$ in shape and phase. The indefinite precession at $x$ = 92.3 becomes apparent with a positive $\delta_0$ implying that one driving mechanism starts dominating with increasing $I_P$. Moreover, the phase appears to be partially inverted for $x$ = 95.3. In Fig. 1c, we display the $I_P$ dependence of normalized $\Delta\theta(t)$ for $x$ = 95.3, where the largest variation in $\delta_0$ occurs.

Fig. 2a shows the $I_P$ dependence of $\delta_0$ for all compositions $x$. Except for $x$ = 95.3, the $\delta_0$ show little changes, while the film with $x$ = 95.3 exhibits the largest change in $\delta_0$ from -94 ° to 1 °. Although π-phase inversion (from -90 ° to 90 °) by a further $I_P$ was expected, it was not realized completely because $I_{P,high}$ already reached approximately 90 % of the burning threshold of the films. Interestingly, the graphs in Fig. 1c indicate that $\delta_0$ changes continuously within one material with $I_P$, signifying the possibility of fine-tuning $\delta_0$. In addition, there is a nonzero background offset of $\delta_0$ in Fig. 2a (pink dashed line). We ascribe this to the latent time required for the dominant driving mechanism of spin precession to become definite through the exchange of thermal energy and angular momentum among the electron, lattice, and spin systems.

We note an interesting additional feature manifesting a shift in the driving mechanism of the spin precession. Fig. 2b presents the $f$ extracted using Eqn. (1) for all $x$ as a function of $I_P$. For $x = 87.0$ and $90.9$, showing positive $\delta_0$, $f$ shows a negative change by -0.5 and -0.3 % as $I_P$ increases, respectively. In contrast, for $x = 95.3$ and $97.5$, showing negative $\delta_0$, $f$ increases by 2.5 and 2.1 %, respectively.

*3.2 Simulation results and comparison with experimental data*

We address that the trends of $\delta_0$ and $f$ as functions of $x$ and $I_P$ are attributed to the change in the relative dominance between the conventional and magnetoelastic effects. The total free energy $E_{tot}$ of a magnetic material is comprised of the magnetocrystalline, demagnetizing, Zeeman, and magnetoelastic energy terms as follows:

$$E_{tot} = K_u(M)\sin^2\theta + \frac{1}{2}\mu_0 \sum_{i=x,y,z} N_i M_i^2(t) - \mu_0 \sum_{i=x,y,z} M_i(t) H_{ext,i} - \frac{3}{2}\lambda_s \sigma_s \cos^2\theta. \qquad (2)$$

Here, $K_u$, $M$, and $N$ are the uniaxial magnetic anisotropy coefficient, magnetization, and demagnetizing factor, respectively. $\sigma_s = 3(1 - v)(1 + v)^{-1} B \eta_{qss}(t)$ is the mechanical stress expressed by Poisson's ratio $v$, bulk modulus $B$, and time-dependent quasi-static strain $\eta_{qss}(t)$ originated from the thermal lattice expansion. $\theta$ is defined as the angle between $\mathbf{M}(t)$ and the $+z$ direction. Generally, in ultrafast photo-induced experiments, the thermal energy increases the temperature and subordinately produces the quasi-static strain owing to lattice expansion. Hence, it is important to distinguish between the pure strain effect and pure thermal effect on the ultrafast dynamics. The first three terms, referred to as the conventional energy term, represent the thermal effect whereas the magnetoelastic energy term describes the pure strain effect. The existence of the quasi-static strain and its applicability to the magnetoelastic energy term have been experimentally demonstrated with ultrafast Sagnac interferometric measurements in Ref [28]. Therefore, without further discussion, it is reasonable to use the magnetoelastic energy term to account for the quasi-static strain effect.

To explain experimental curves, we performed the simulation of the spin precessions for all $x$ using the LLG equation incorporating the time-dependent temperatures of the electron, lattice, and spin systems.

$$\frac{d\mathbf{M}(t)}{dt} = -\gamma_s \mu_0 (\mathbf{M}(t) \times \mathbf{H}_{\text{eff}}(t)) + \frac{\alpha_d}{M}(\mathbf{M}(t) \times \frac{d\mathbf{M}(t)}{dt}), \quad (3)$$

where $\gamma_s$ and $M$ are the gyromagnetic ratio and magnetization, and the effective magnetic field vector defined as $\mathbf{H}_{\text{eff}} = -\partial E_{\text{tot}}/\mu_0 \partial \mathbf{M}$. The phenomenological damping parameter $\alpha_d$ is assumed to be 0.015 for all $x$. In order to avoid misinterpretation as much as possible, unknown values of the magnetic parameters used in the simulation were extracted by linear interpolations from the values for $Ni_{80}Fe_{20}$ (Py) and Ni. However, if the variation in a parameter between two compositions was insignificant or its effect on the spin precession dynamics was negligible, Ni data were used instead. For $\gamma_s$, the linear relation between $\gamma_s = 1.70 \times 10^{11}$ (for $Ni_{90}Fe_{10}$) [30] and $1.61 \times 10^{11}$ rad/s·T (for Ni) [31] was employed. As common values used regardless of $x$, we set the uniaxial magnetic anisotropy coefficient $K_u \sim 0$, demagnetizing factors $N_x = N_y = 0$, $N_z = 1$, and $v = 0.31$, and bulk modulus $B = 180$ GPa [32].

The time-dependent profiles of $\mathbf{M}(t)$ and $\eta_{\text{qss}}(t)$ required for solving the LLG equation are obtained using the three-temperatures model described as follows:

$$C_i(T_i)\frac{\partial T_i}{\partial t} = \delta_{\text{ie}}\left[\frac{\partial}{\partial z}\left(\kappa \frac{\partial T_i}{\partial z}\right) + P(z,t)\right] - g_{ij}(T_i - T_j), \quad (4)$$

where $P(z, t)$ represents the laser source term, and $i, j = e, l, s$ denote electron, lattice, and spin sub-systems, respectively. $C_i$ is the heat capacity of the system $i$, and the respective values were used as follows: $\gamma_e$ (= $C_e/T_e$) = $6.25 \times 10^2$ (Py) [33] and $1.23 \times 10^3$ Jm$^{-3}$K$^{-2}$ (Ni) [34], $C_l = 3.7 \times 10^6 + 500 T_l$ (Ni data used for all $x$) [35], and $C_s = m_1(1-T_s/T_c)^{0.1} + m_2$, where $m_1 = -2.15 \times 10^6$ and $m_2 = 2.38 \times 10^6$ Jm$^{-3}$K$^{-1}$ under the condition $T_s < T_c$ [36]. The composition-dependent Curie temperature $T_c$ was extracted from the data in Ref [37]. For the Curie-Weiss curve, $M(T) = M_0(1-1.058(T/T_c)^\alpha)^\xi$ was extracted by fitting the data in Ref [38], here $\alpha = 2.80$ and $\xi = 0.42$. The saturation magnetization ($M_s$) at room temperature was experimentally measured using a vibrating sample magnetometer. The thermal conductivity $\kappa$ was extracted from the data in Ref [39]. The following values were used for the dynamic heat coupling coefficients: $g_{es} = 3.0 \times 10^{17}$ (for all $x$), $g_{sl} = 2.0 \times 10^{17}$ (for all $x$), and $g_{el}(x, T) = ((-1.65 \times 10^4 - 54\, T_e + 1.97 \times 10^{-2}\, T_e^2)x + 3 \times 10^6 + 4.32 \times 10^3\, T_e - 1.6\, T_e^2)10^{12}$ Wm$^{-3}$K$^{-1}$. These were extracted by fitting the data for the range 300 K $\leq T \leq$ 1000 K in Ref [35]. We summarized the parameter values for the five compositions used in the simulations in Table 1. Based on these parameter values, by putting the solution of a lattice temperature $T_l(t)$ from Eqn. (4) into one-dimensional wave equation, the lattice displacement and corresponding time-dependent $\eta_{qss}(t)$ are easily obtained [28]. Then, the magnetization dynamics can be finally simulated by solving LLG equation. Here, for $\lambda_s$ which is sensitively dependent on other factors, such as thickness and fabrication process [40,41], we chose the values for each $x$ that best reproduced the experimental data in Fig. 2 rather than adopting the literature values.

Fig. 3a and 3b show the simulation results for the magnetization dynamics $\Delta M_z(t)/M_z$ obtained by solving Eqns. (2) – (4) representatively at the lowest ($I_p = I_0$) (Fig. 3a) and highest pump fluences ($I_p = 13I_0$) (Fig. 3b) for all $x$'s. The highest pump fluence in the simulation was chosen to reach $T_s \sim 0.99 T_c$ for $x = 97.5$, where $T_c = 659$ K. In Fig. 3a, the phase inversion between $x = 92.3$ and $95.3$ is reproduced well, although the curve for $x = 92.3$ exhibits a more definite precession compared to the experimental data. Importantly, the curves for $x = 95.3$ in Fig. 3a and 3b demonstrates the signature of the phase transition. In Fig. 3c, we present the simulation results $\Delta M_z(t)/M_z$ for $x = 95.3$. The simulation reproduced excellently the continuous change of the phase matching with the experimental data in Fig. 2a. We emphasize that only one free parameter ($\lambda_s$) for each $x$ is used to obtain all simulation results. In Fig. 3d, we plot $\lambda_s$ used in the simulation (red circles) and $\langle \lambda_s \rangle = (2\lambda_{100} + 3\lambda_{111})/5$ for a polycrystalline structure (blue line) retrieved from the bulk sample in the literature [37]. The graphs show that the $\lambda_s$ data points do not lie exactly on the curve of $\langle \lambda_s \rangle$. In general, the physical parameters involved in the magnetoelasticity of a thin film, such as $B$, $v$, and $\lambda_s$ would be sensitive to a thickness, for which one of the main reasons is the interfacial strain between the film and substrate [42,43]. Here, since we set $B$ and $v$ to constants, their possible deviation from bulk values can melt into $\lambda_s$ and may explain the difference between $\lambda_s$ and $\langle \lambda_s \rangle$ despite qualitative agreement. In Fig. 3e and 3f, we summarize the simulation results for both $\delta_0$ and $f$ as functions of $I_p$ for all $x$, respectively. These results fairly reproduce the experimental data in the viewpoints of both the large change in $\delta_0$ for $x = 95.3$ and $\Delta f < 0$ ($> 0$) for $x < 92.3$ ($> 92.3$) as functions of $I_p$.

For a case of inplane magnetic anisotropic material under external magnetic field angle $\varphi < 90°$, a sudden decrease in $M(t)$ by photoexcitation induces a rotation of $\mathbf{H}_{\text{eff}}(t)$ toward the $z$-direction, primarily owing to the decrease of the demagnetizing energy. This results in both $\delta_0 > 0$ and the decrease of the spin precession frequency ($\Delta f < 0$). This phenomenon is expected to continue as long as the increase of the magnetoelastic field does not exceed the decrease of the conventional effective field, corresponding to materials for $x < 92.3$. On the other hand, when the magnetoelastic field effect dominates the conventional one for a certain condition such as a sufficiently large value of a lattice expansion or a large negative value of $\lambda_s$, $\mathbf{H}_{\text{eff}}(t)$ reorients to the $xy$ plane, leading to a negative $\delta_0$. At the same time, an increase of the magnetoelastic field exceeds a decrease of the conventional effective field, resulting in $\Delta f > 0$. This corresponds to the case of $x \geq 97.5$. Therefore, the effective fields from the two terms are expected to compete within a certain range of $\lambda_s$, producing a large variation in $\delta_0$ like a case of $x = 95.3$.

*3.3 Effective magnetic field calculations and key material parameters for simple analysis*

Normally, it is not easy task to consistently analyze all experimental results for different materials as the properties of all materials change when their compositions vary. Therefore, it is meaningful to provide a simple way for understanding the large variation in $\delta_0$ with $I_p$ for a fixed concentration, $x$. Because spin precession in a polycrystalline structure is initially triggered by a change in the azimuthal component of the total effective magnetic field, it is instructive to calculate the $\theta$ component of the effective magnetic fields for the respective energy terms.

$$H_{d,\theta}(t) = M(t)\sin\theta(t)\cos\theta(t),$$
$$H_{Z,\theta}(t) = H_{\text{ext}}\sin(\varphi - \theta(t)), \quad (5)$$
$$H_{m,\theta}(t) = -\frac{9}{\mu_0 M(t)}\frac{1-\nu}{1+\nu}B\lambda_s\eta_{\text{qss}}(t)\sin\theta(t)\cos\theta(t).$$

Here, $H_{d,\theta}$, $H_{Z,\theta}$, and $H_{m,\theta}$ represent the $\theta$ component of the effective field with respect to the demagnetizing, Zeeman, and magnetoelastic energy terms, respectively. The pump-induced change in $H_{d,\theta}$ was calculated as $\Delta H_{d,\theta}(t) = M(t)\sin\theta(t)\cos\theta(t) - M_s\sin\theta_0\cos\theta_0$, where '0' in $\theta_0$ denotes the equilibrium value before the arrival of the pump. Since the photo-excitation induces $\Delta M(t) < 0$ and $\Delta\theta < 0$ because of the loss of demagnetizing energy, the signs of $\Delta H_{d,\theta}(t)$, $\Delta H_{Z,\theta}(t)$, and $\Delta H_{m,\theta}(t)$ become negative, positive, and positive, respectively. Of these, $\Delta H_{Z,\theta}(t)$, which does not contain $M(t)$, has the negligible contribution to $\Delta H_{\text{eff}}(t)$, and eventually $\Delta H_{d,\theta}(t)$ and $\Delta H_{m,\theta}(t)$ mainly competes depending on the composition $x$.

Fig. 4a–4c shows a relative comparison of $-\Delta H_{d,\theta}$ (red) and $\Delta H_{m,\theta}$ (blue) at two extreme $I_p$ values for $x = 87.0$, $95.3$, and $97.5$, respectively. In this simulation, as $\Delta\theta$ has only a few degrees, we set $\theta(t) = \theta_0$ for convenience. For all $x$, $-\Delta H_{d,\theta}$ is enhanced compared to $-\Delta H_{m,\theta}$ as $I_p$ increases mainly due to the nonlinear temperature dependence of $M(t)$ according to the Curie-Weiss law. In Fig. 4a, $-\Delta H_{d,\theta}$ is larger than $\Delta H_{m,\theta}$ at $I_p = I_0$, and the difference between $-\Delta H_{d,\theta}$ and $\Delta H_{m,\theta}$ is enhanced with increasing $I_p$. This causes the total effective field to rotate toward the $+z$ direction, further maintaining $\delta_0 > 0$.

In Fig. 4c, $\Delta H_{m,\theta}$ (blue) is much larger than $-\Delta H_{d,\theta}$ (red) at $I_p = I_0$, yielding the total effective field toward $xy$ plane with $\delta_0 < 0$. Although the difference between $-\Delta H_{d,\theta}$ and $\Delta H_{m,\theta}$ is reduced at $I_p = 13I_0$, the condition $-\Delta H_{d,\theta} < \Delta H_{m,\theta}$ still holds. Hence, it is reasonable to think that $\delta_0$ starts from a negative value and slightly goes to the positive side (but remains negative) with increasing $I_p$ as shown in Fig. 2a and Fig. 3e. Importantly, in Fig. 4b, starting from the condition that $-\Delta H_{d,\theta}$ is slightly smaller than $\Delta H_{m,\theta}$ at $I_p = I_0$, $-\Delta H_{d,\theta}$ and $\Delta H_{m,\theta}$ become comparable at $I_p = 13I_0$. This leads to a large transition of $\delta_0$ as shown by the pink curve in Fig. 2a. One may think of the total suppression of precession by fine-tuning of $I_p$. However, the different profiles of the two effective fields make it obscure to cancel out completely over the entire timescale.

We explored the relationship between the parameters in detail for a clear understanding of the phase of spin precession. The phase is initially determined by various material parameters, such as $g_{ij}$, $C_i$, $T_c$, $M_s$, $B$, $\lambda_s$, $\beta$ (thermal expansion coefficient). While $g_{ij}$ and $C_i$ are the main factors in determining the temperature profiles of sub-systems, other parameters ($T_c$, $M_s$, $B$, $\lambda_s$, and $\beta$) participate in the spin dynamics after the temperature profiles are set. Because $B$ and $\beta$ are less sensitive to $x$, it is instructive to examine the relation among $T_c$, $M_s$, and $\lambda_s$ to derive a useful picture. The transition of $\delta_0$ is closely linked to the relative strength between $-\Delta H_{d,\theta}$ and $\Delta H_{m,\theta}$ with respect to the temperature change $\Delta T$. The ratio of the two effective fields - $\Delta H_{d,\theta}/\Delta H_{m,\theta}$ can be written as $(M_s(t<0)+\Delta M)(\Delta M - H_{ext})/B\lambda_s\eta_{qss}$. Simply setting $\eta_{qss} \sim \beta\Delta T$, the relationship is simplified to $\Delta M^2/\lambda_s\Delta T$.

In Fig. 4d–4f, we present the temperature dependence of $-\Delta H_{d,\theta}$ (orange surface) and $\Delta H_{m,\theta}$ (blue surface) for the three cases varying the parameter values $T_c$, $M_s$, and $\lambda_s$, respectively. To investigate the effect of the parameters independently, when we varied the value of one parameter from $x = 100 - 80$, while the other two parameters were assigned to the mean values of Ni and Py for convenience. For examples, $\bar{T}_c = (T_{c,Ni}+T_{c,Py})/2$, $\bar{M}_s = (M_{s,Ni}+M_{s,Py})/2$, and $\bar{\lambda}_s = (\lambda_{s,Ni}+\lambda_{s,Py})/2$. Importantly, as a common feature for Fig. 4d-4f, at a given parameter value, while $\Delta H_{m,\theta}$ increases almost linearly with increasing $T$ (or $I_p$), $-\Delta H_{d,\theta}$ grows rapidly and ends up dominating $\Delta H_{m,\theta}$ primarily because of the nonlinear behavior of $\Delta M$. The nonlinearity of $\Delta M$ is intensified when $T$ reaches $T_c$. Then, let us check the parameter dependence at a fixed $T$. From the lower panel of Fig. 4d (graphs intersected by a semitransparent plane), we see that $\Delta H_{m,\theta}$ overtakes $-\Delta H_{d,\theta}$ when $T_c$ varies from $T_{c,Ni} = 631$ K to $T_{c,Py} = 853$ K. For a given $\Delta T$, $\Delta M$ is expected to decreases as $T_c$ increases, according to the Curie-Weiss law. This causes $\Delta M^2/\lambda_s\Delta T$ to decrease, indicating that the demagnetizing effect becomes weaker than the magnetoelastic effect. The graphs in Fig. 4e and its lower panel show that when $M_s$ changes from $\mu_0 M_{s,Ni} = 0.58$ T to $\mu_0 M_{s,Py} = 1.02$ T, $-\Delta H_{d,\theta}$ monotonically increases and begins to cross over $\Delta H_{m,\theta}$. This is because $\Delta M$ is proportional to $M_s$ at a given $\Delta T$, therefore $\Delta M^2/\lambda_s\Delta T$ increases. Lastly, when $\lambda_s$ changes from $\lambda_{s,Ni} = -40\,\mu$ to $\lambda_{s,Py} = 0$ as shown in Fig. 4f and its lower panel, it is evident that $\Delta M^2/\lambda_s\Delta T$ increases, leading to the stronger demagnetizing effect.

We note that in a real experimental situation, a different composition concurrently changes all parameter values as well as $T_c$, $M_s$, and $\lambda_s$, such as static values – bulk modulus ($B$), heat capacities ($C_l$ and $C_s$) and dynamic values - heat coupling coefficients ($g_{es}$ and $g_{sl}$), which are assumed to be independent on our composition range. Hence, this makes systematic understanding of spin dynamics difficult. Nevertheless, the way of simple and comprehensive analysis based on independent variation of three key parameters helps identify the hidden driving mechanism of spin precession.

## 4. Summary

To demonstrate the key role of magnetoelasticity, we performed time-resolved magneto-optical Kerr effect measurements of spin precession for a series of $Ni_xFe_{100-x}$ thin films ($x =$ 87–97.5). First, we successfully observed the phase inversion of the spin precession according to the atomic composition. Second, the phase of the spin precession showed a dramatic transition with a sign change at the composition of $x = 95.3$ as a function of the pump fluence. Finally, the frequency change of the spin precession with the pump fluence exhibited a sign dependence on $x$. Through the simulations using the three-temperatures model and LLG equation, including the magnetoelasticity, we can fairly reproduce all the experimental features mentioned above. The transition between the magnetoelastic and demagnetizing-Zeeman fields was described by the interplay among the key parameters $T_c$, $M_s$, and $\lambda_s$, that determined the driving mechanism of spin precession dynamics. Our extensive work, performed by varying material properties, clearly demonstrates that the magnetoelastic effect is essential for a complete understanding of spin dynamics and at the same time, proposes the new method of controlling the spin-orbit coupling strength through interfacial strain modification at inversion symmetry broken heterostructures for opto-spintronic devices [44].

**Declaration of Competing Interest**

The authors declare no conflict of interest.

**Acknowledgements**

J.-W.K. conceived the experiment. S.Y. and J.-I.H. fabricated NiFe alloy films. Y.S. and S.Y. conducted TR-MOKE experiment. J.-W.K. performed data analysis and simulation with the help of Y.S. J.-W.K. and J.-I.H. wrote the manuscript. All authors discussed the results and commented on the manuscript. Y. Shin and S. Yoon contributed equally to this work.

**Funding**

This work was supported by Basic Science Research Program through the National Research Foundation of Korea (NRF) funded by the Ministry of Education (2020R1I1A1A01075040, 2021M3H4A6A02045430, 2022R1I1A3072023) and by the MSIT (2020R1A2C2005932, 2021R1A4A1031920).

**Figure captions**

**Fig. 1.** Ultrafast spin dynamics with various pump fluences for $Ni_xFe_{100-x}$ films. Differential polar Kerr rotation $\Delta\theta(t)$ measured at the (a) lowest pump fluence $I_{P,low} = 0.21$ mJ/cm$^2$ and (b) highest pump fluence $I_{P,high} = 4.6$ mJ/cm$^2$. (c) Dependence of the normalized $\Delta\theta(t)$ on the pump fluence ($I_P$) for $x = 95.3$. The solid lines on the data for $x = 97.5$ in (a) and (b) represent curves fitted by the damped sinusoidal function (Eqn. (1)).

**Fig. 2.** Pump fluence dependence of the phase and the frequency extracted from the fitting. $I_P$ dependence data of (a) the spin precession phase ($\delta_0$) and (b) the spin precession frequency ($f$) extracted using the damped sinusoidal function for all atomic compositions $x$.

**Fig. 3.** Simulation results for ultrafast spin precession. Magnetization dynamics $\Delta M_z(t)/M_z$ at the (a) lowest pump fluence ($I_p = I_0$) and (b) highest pump fluence ($I_p = 13I_0$) for all $x$. (c) $I_P$ dependence of the magnetization precession curves $\Delta M_z(t)/M_z$ for $x = 95.3$. (d) Comparison of magnetostriction values $\lambda_s$ between the earlier work and ours. The red circles denote the data that best fit the experimental curves. The blue line represents the data extracted from the Ref [37]. The bra-ket symbol stands for the average along crystalline orientations [100] and [111]. Reproduction of (e) $\delta_0$ and (f) $f$ as functions of $I_p$ for all $x$.

**Fig. 4.** Effective magnetic field calculations from respective energy terms for understanding the driving mechanism of spin dynamics. Simulated time-dependent azimuthal components of the demagnetizing ($\Delta H_{d,\theta}$: red) and magnetoelastic fields ($\Delta H_{m,\theta}$: blue) for (a) $x = 87.0$, (b) $x = 95.3$, and (c) $x = 97.5$, respectively. The solid (dashed) lines in the upper (lower) panel in each figure represent the data simulated for $I_p = 13I_0$ ($I_p = I_0$). Temperature-dependent three dimensional plots of $\Delta H_{d,\theta}$ (orange) and $\Delta H_{m,\theta}$ (blue) varying (d) the Curie temperature $T_c$, (e) the magnetization $M_0$, and (f) the magnetostriction $\lambda_s$ corresponding to values from $x = 80$ to 100, respectively. The graphs at the lower panel of each figure (d, e, f) display the $\Delta H_{d,\theta}$ (orange) and $\Delta H_{m,\theta}$ (blue) curves where each semi-transparent plane intersects.

**Table 1** Summary of the parameter values used in the simulations.

**Figures**

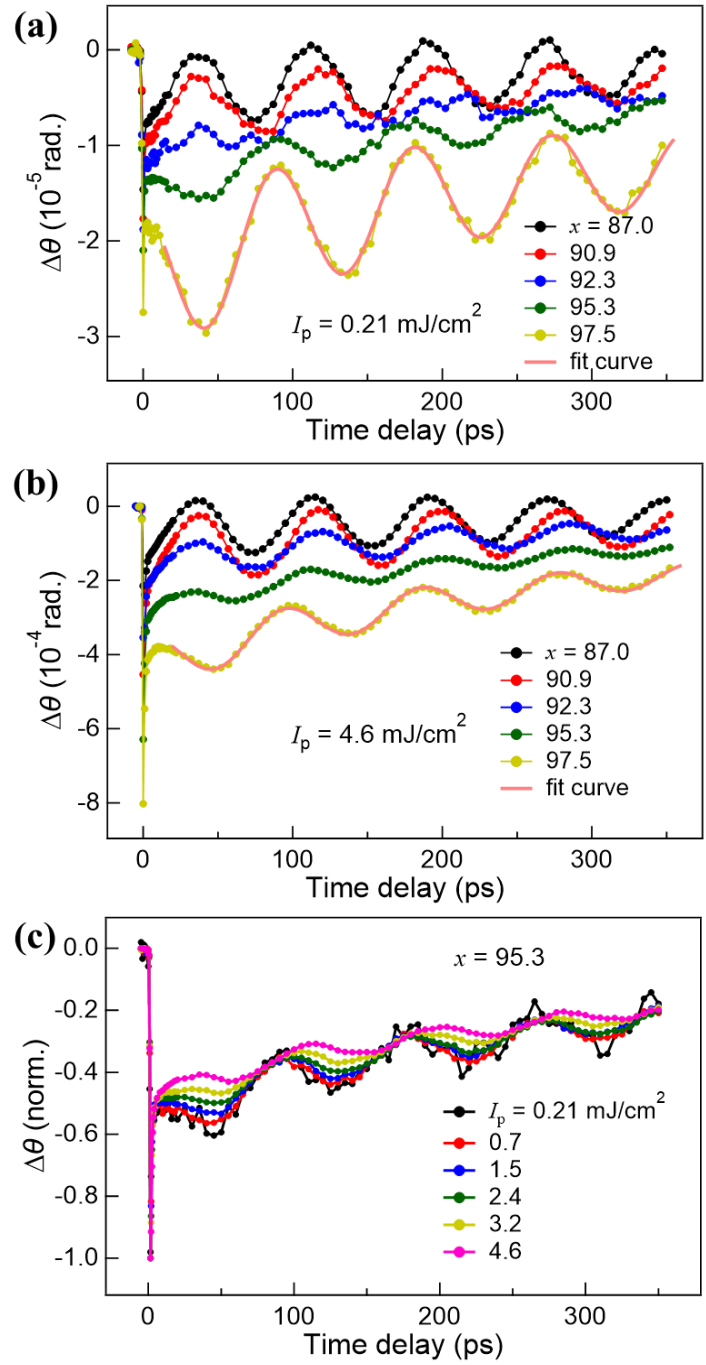

Figure 1

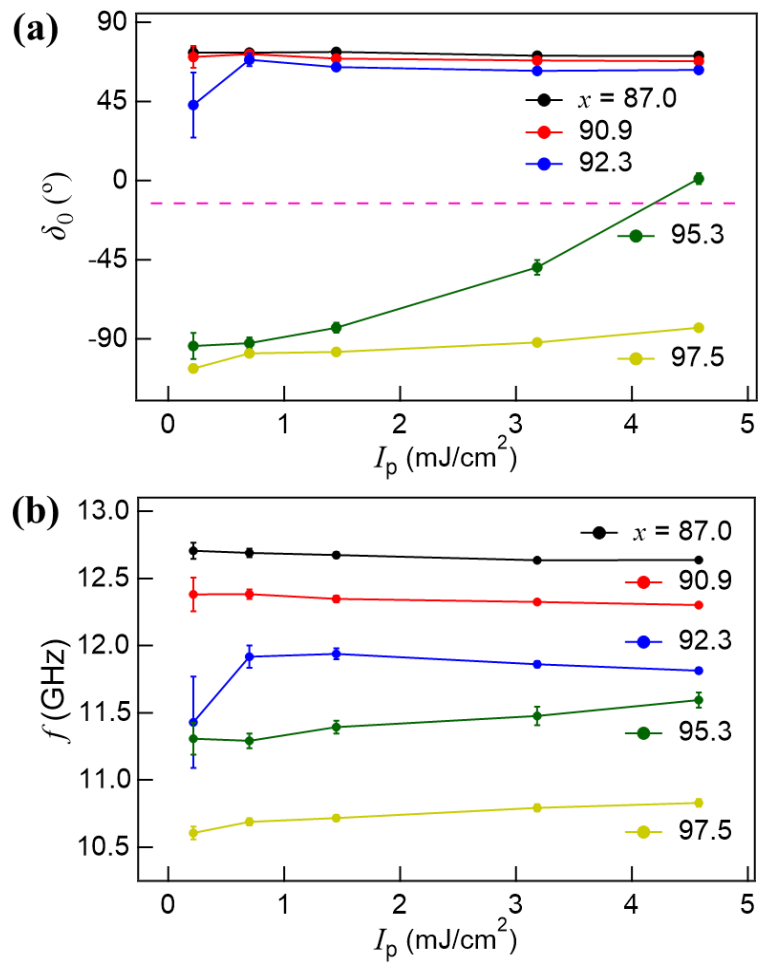

Figure 2

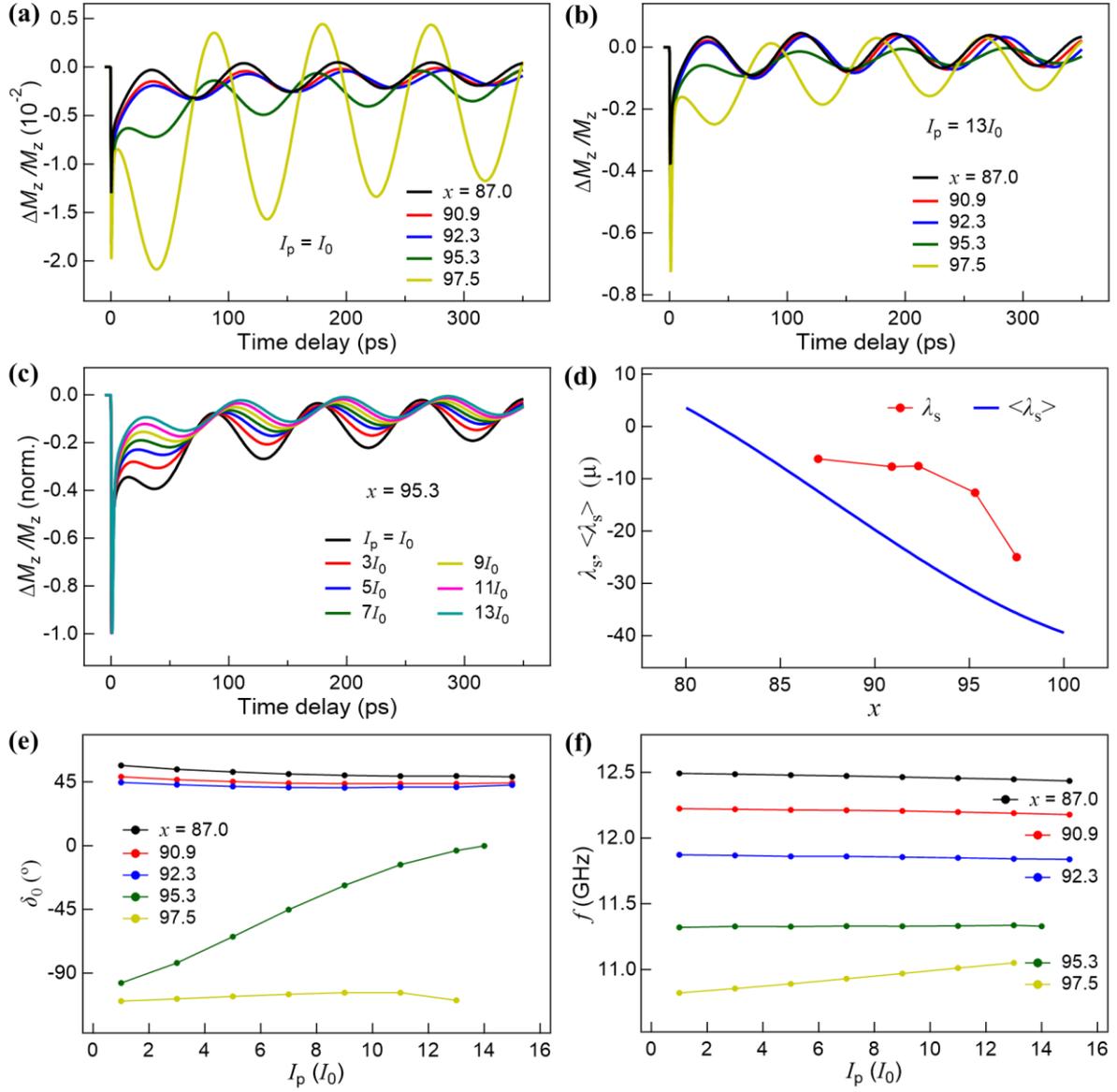

Figure 3

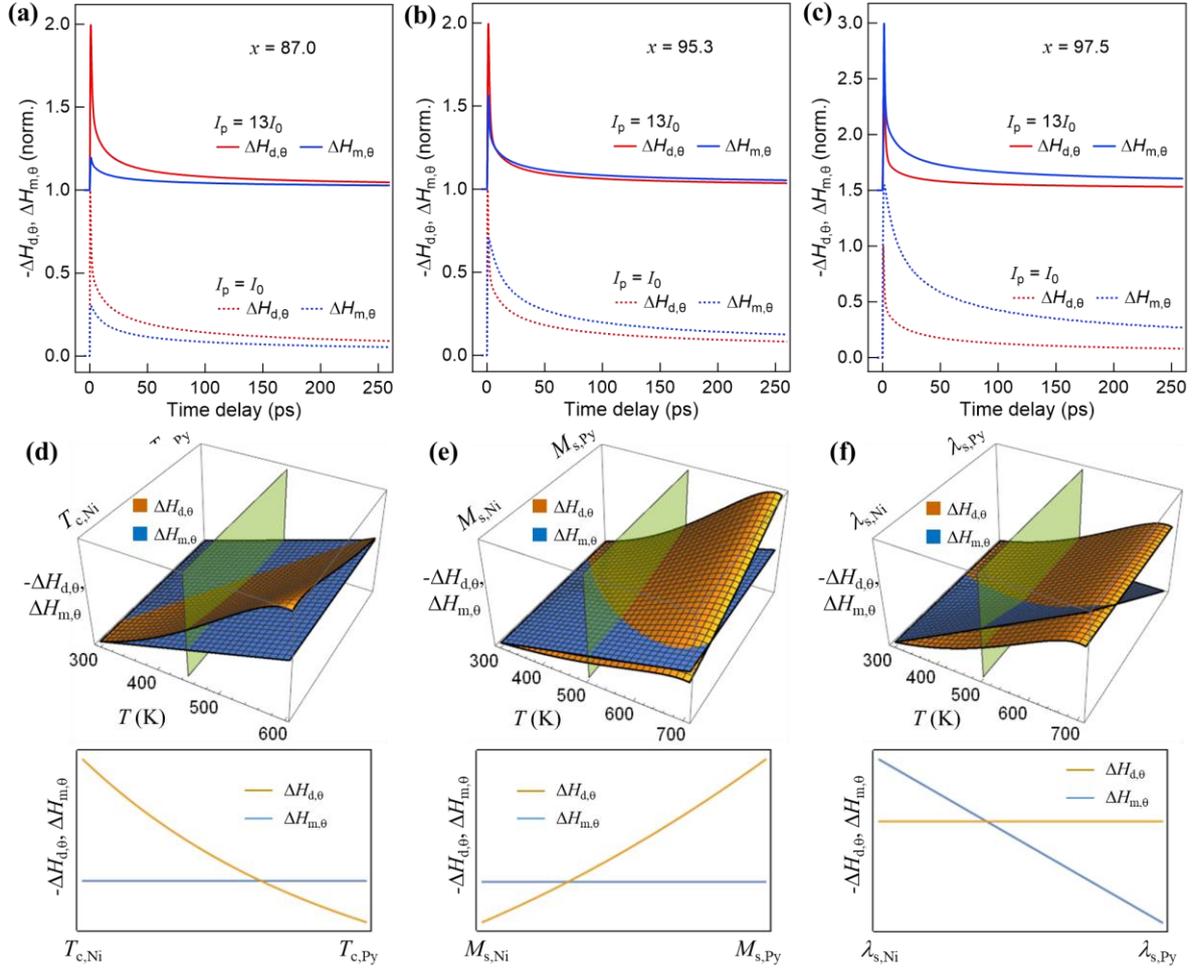

Figure 4

**Tables**

| | **x = 87.0** | **90.9** | **92.3** | **95.3** | **97.5** |
|---|---|---|---|---|---|
| $\gamma_s$ [$10^{11}$ s$^{-1}$T$^{-1}$] | 1.667 | 1.650 | 1.644 | 1.631 | 1.621 |
| $\mu_0 M_s$ [T] | 0.898 | 0.877 | 0.824 | 0.743 | 0.670 |
| $\gamma_e$ [Jm$^{-3}$K$^{-2}$] | 836.8 | 954.7 | 997.1 | 1087.8 | 1154.4 |
| $C_l$ [Jm$^{-3}$K$^{-1}$] | \multicolumn{5}{c}{$3.7\times10^6 + 500\times T_l$} |
| $C_s$ [Jm$^{-3}$K$^{-1}$] | \multicolumn{5}{c}{$(-2.15(1-T_s/T_c)^{0.1} + 2.38)\times10^6$} |
| $g_{el}$ [$10^{12}$ Wm$^{-3}$K$^{-1}$] | $a + bT_e + cT_e^2$ ||||| 
| | $a = 1.56\times10^6$ $b = -378.0$ $c = 0.114$ | $1.50\times10^6$ $-588.6$ $0.191$ | $1.48\times10^6$ $-664.2$ $0.218$ | $1.43\times10^6$ $-826.2$ $0.277$ | $1.39\times10^6$ $-945.0$ $0.321$ |
| $g_{es}$ [Wm$^{-3}$K$^{-1}$] | \multicolumn{5}{c}{$3.0\times10^{17}$} |
| $g_{sl}$ [Wm$^{-3}$K$^{-1}$] | \multicolumn{5}{c}{$2.0\times10^{17}$} |
| $\kappa$ [Wm$^{-1}$K$^{-1}$] | 49.9 | 53.1 | 54.9 | 61.5 | 71.1 |
| $T_c$ [K] | 769.2 | 728.1 | 712.6 | 681.3 | 657.9 |
| $B$ [GPa] | \multicolumn{5}{c}{180} |
| $v$ | \multicolumn{5}{c}{0.31} |
| $\beta$ [µ/K] | \multicolumn{5}{c}{13} |

Table 1